# Photonuclear reactions on the stable isotopes of selenium at bremsstrahlung end-point energies of 10-23 MeV

F.A. Rasulova[a,b, *], N.V. Aksenov[a], S.I. Alekseev[a], R.A. Aliev[c,d] S.S. Belyshev[e,f], I. Chuprakov[a,e], N.Yu. Fursova[e,f], A.S. Madumarov[a], J.H. Khushvaktov[a,b], A.A. Kuznetsov[e,f], B.S. Yuldashev[a,b]

[a] Joint Institute for Nuclear Research, Dubna, Russia
[b] Institute of Nuclear Physics of the Academy of Sciences of the Republic of Uzbekistan, Tashkent, Uzbekistan
[c] Faculty of Chemistry of Lomonosov Moscow State University, Moscow, Russia
[d] National Research Center "Kurchatov Institute", Moscow, Russia
[e] Skobeltsyn Institute of Nuclear Physics of Lomonosov Moscow State University, Moscow, Russia
[f] Faculty of Physics of Lomonosov Moscow State University, Moscow, Russia
[e] Institute of Nuclear Physics, Almaty, Republic of Kazakhstan
**rasulova@jinr.ru*

**Abstract**

The experiments were performed at bremsstrahlung end-point energies of 10-23 MeV with the beam from the MT-25 microtron with the use of the γ-activation technique. The experimental values of relative yields were compared with theoretical results obtained on the basis of TALYS with the standard parameters and the combined model of photonucleon reactions. Including isospin splitting in the combined model of photonucleon reactions allows to describe experimental data on reactions with proton escape in energies range from 10 to 23 MeV. Therefore, taking into account isospin splitting is necessary for a correct description of the decay of the GDR.

## 1. Introduction

Photonuclear reactions play an important role in basic and applied nuclear physics research [1-4]. The development of both nuclear physics research and applications based on photonuclear reactions is closely related to the development of attendant photon sources and measuring instruments. Photoneutron reactions on isotopes of natural mixture of selenium have been well studied using bremsstrahlung γ-radiation [5-14], positron annihilation in flight [15] and Compton backscattering of laser beam photons [16-20]. Cross sections of photonuclear reactions at selenium isotopes in the region of energy giant dipole resonance were measured in several works. In [5] on bremsstrahlung the cross section for reactions σ(γ, *n*) on all isotopes of natural selenium were measured up to energy 25 MeV. In [6], the same method was used to measure the cross sections for the reactions $^{78}$Se(γ, *n*)$^{77m}$Se, $^{80}$Se(γ, $n)^{79m}$Se and $^{82}$Se($\gamma$, $n)^{81m}$Se. In [16, 17, 18, 19, 20], on a beam of quasi-monoenergetic photons obtained as a result of Compton backscattering, the cross sections for reactions $\sigma(\gamma, n)$ were measured in the energy range from threshold to 14.6 MeV on $^{76,77,78,80}$Se isotopes. In [15] on a beam of quasi-monoenergetic photons, the reaction cross section $\sigma(\gamma, 2n)$ and the sum of the reaction cross sections $\sigma(\gamma, n)$ + $\sigma(\gamma, 1n1p)$ were measured in the energy range from threshold to 30 MeV on $^{76,78,80,82}$Se isotopes.

Experimental data on the cross sections of photoproton reactions on selenium isotopes are not available in the literature. With the exception of our previous works [13,14], wherein multiparticle reactions on natural selenium were studied using end-point energies ranging from 20 to 80 MeV, all other previous research investigated only photoneutron reactions on the stable isotopes of selenium. Thus, to obtain more information regarding reactions with a higher degree of complexity, this study investigated photonucleon emission reactions on natural selenium target nuclei, expressed as $^{nat}$Se($\gamma$, 1$n$) and $^{nat}$Se($\gamma$, 1$p$), using bremsstrahlung end-point energies of 10 to 23 MeV.

Photonuclear reactions are the main mechanism behind the formation of bypassed nuclei in the process of nucleosynthesis. The abundance of the lightest $p$-nucleus $^{74}$Se can be described satisfactorily [21]. One of the aims of this work was to measure the relative yields of reactions on mixtures of natural isotopes of selenium that result in the formation and decay of $^{74}$Se. Experimentally obtained results are compared with the results of calculations based on the TALYS-1.96 [22] with the standart parameters and combined model of photonucleon reactions (CMPR) [23]. In addition, photoproton reaction products are potential medical isotopes, namely, $^{76}$As [24] and $^{77}$As [25,26]; this means that studying the reaction cross sections is useful for both research and application purposes.

## 2. Experimental Set-up and procedures

This work was performed with the output electron beam of the MT-25 microtron [27]. The electron energies were in range of 10-23 MeV with an energy step of 1 MeV. To produce gamma radiation a radiator target made of tungsten, which is a common convertor material, was used. The tungsten target was thick enough (3 mm) to maximize the number of photons in the energy range of the giant dipole resonance which dominates the photonuclear cross section from the nucleon separation threshold to 20–30 MeV. To remove the remaining electrons from the bremsstrahlung beam, an aluminum absorber 30 mm thick was placed behind the tungsten converter [28]. The target of natural selenium was at a distance of 1 cm from the converter.

In the experiments natural selenium samples in metallic form were irradiated with a flux of bremsstrahlung, which was formed in a tungsten converter. Changes in beam current were measured using a calibrated ionization chamber in the beam and Faraday cup and recorded in a web-accessible database for use during analysis using an analog-to-digital converter card and LabView software.The electrical charge collected on the target was also digitized and used to measure the beam current in addition to the ionization chamber and Faraday cup. The main parameters of experiments are given in the Table 1. After irradiation, when radiation levels in the experimental hall become safe the targets were transferred to a separate measuring room, where the induced activity in the irradiated target was measured. We used High Purity Germanium γ-detector with resolution of 16 keV at 1332 keV in combination with standard measurement electronics and a 16K ADC/MCA (Multiport II Multichannel Analyzer, CANBERRA). The energy and efficiency calibrations of the HPGe detector were carried out using the standard gamma-ray sources. The procedure gamma-activation measurements used in this work was described in detail in [28-30].

The time from the end of irradiation to the start of measurement (cooling time) was in range from 10 to 15 min. For each sample, the spectra were measured at several times during an overall period of 0.5, 1, 2, 6, 12 and 24 hours. Typical γ-ray spectra of the reaction products produced from the $^{nat}Se$ are shown in Fig. 1. The sample irradiated with bremsstrahlung radiation with end-point energy of 23 MeV.

Table 1. The main parameters of experiments

| Energy of electrons, MeV | Mass of selenium target, mg | Electron beam pulse current, μA | Integral number of electrons incident on the tungsten converter, $\times 10^{16}$ | Irradiation time, min |
|---|---|---|---|---|
| 10 | 722.66 | 19 | 42.75 ± 4.28 | 60 |
| 11 | 693.02 | 10 | 45.0 ± 4.5 | 120 |
| 12 | 707.09 | 10.5 | 47.25 ± 4.72 | 120 |
| 13 | 741.20 | 10 | 22.5 ± 2.25 | 60 |
| 14 | 682.06 | 10 | 11.25 ± 1.12 | 30 |
| 15 | 712.51 | 10 | 4.875 ± 0.487 | 13 |
| 16 | 702.77 | 10 | 1.875 ± 0.187 | 5 |
| 17 | 233.48 | 5 | 1.875 ± 0.187 | 10 |
| 18 | 117.83 | 5 | 1.875 ± 0.187 | 10 |
| 19 | 89.26 | 5 | 1.875 ± 0.187 | 10 |
| 20 | 78.35 | 3 | 2.25 ± 0.225 | 20 |
| 21 | 78.36 | 3 | 2.25 ± 0.225 | 20 |
| 22 | 73.48 | 5 | 1.875 ± 0.187 | 10 |
| 23 | 70.22 | 5 | 1.875 ± 0.187 | 10 |

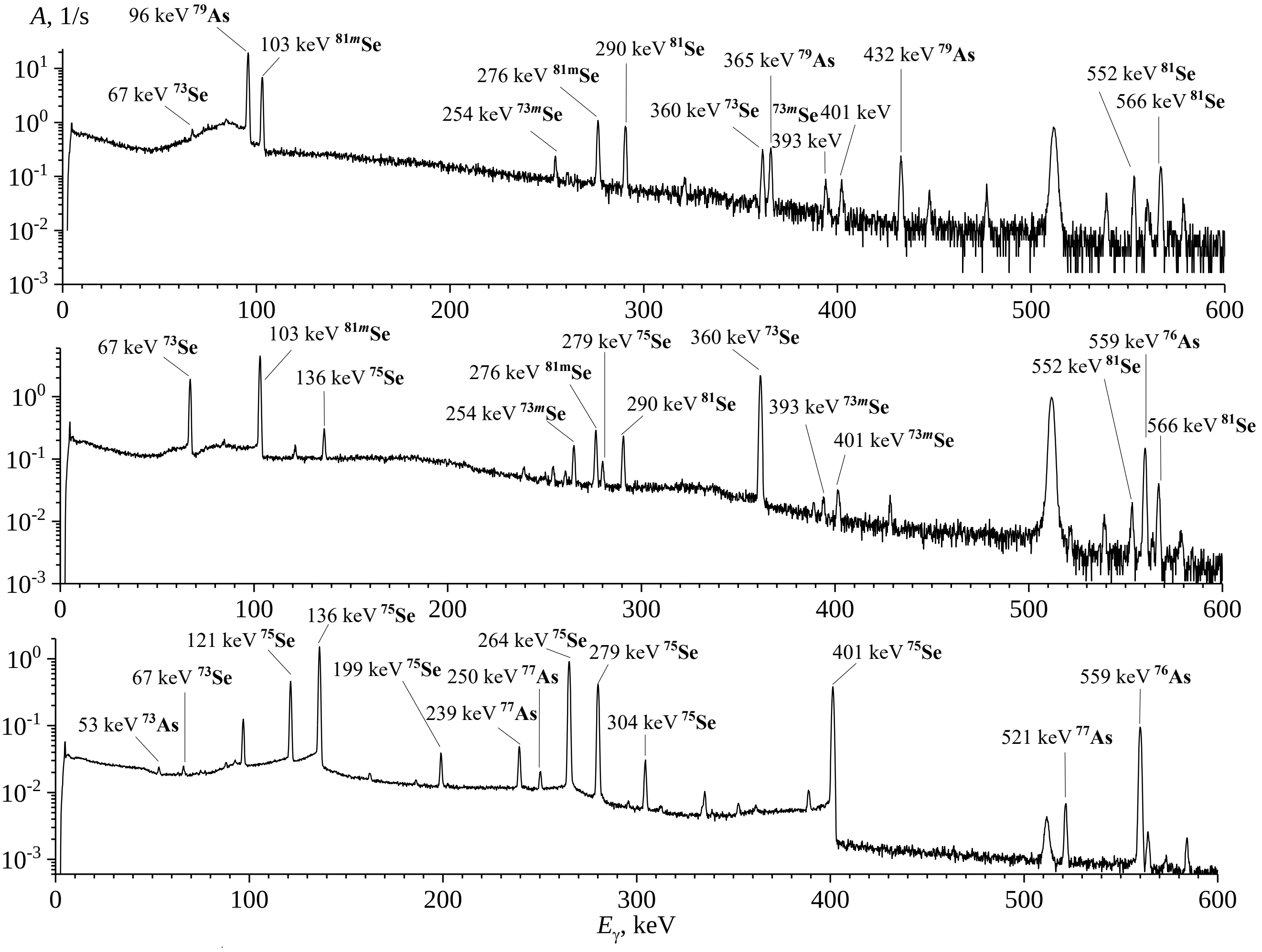

Figure 1. The spectra of residual activity of irradiated sample from a natural mixture of selenium isotopes (top-to-bottom) 10 min, 3 h, and 4 days after irradiation. The spectra measurement duration was 10 min, 1 h, and 1 day, respectively. The bremsstrahlung end-point energy used for the irradiation was 23 MeV

The gamma-ray spectra were processed using the DEIMOS32 code [31], which fits the count area of the full-energy peaks with the Gaussian function. The identification of the processed peaks was based on the gamma-ray energy and intensity, and the half-life of the generated residual nuclei. The radionuclides produced were identified based on their characteristic γ-ray energies and half-lives. The main γ-ray energies and intensities used to determine the yield of the reaction products are given in Table 2. The nuclear data given in columns 4-5 of Table 2 are taken from Ref. [32].

Table 2. Spectroscopic data from ref. [32] for the product nuclei from the photonuclear reactions on the stable isotopes of selenium

| Reaction product | Reactions | $E_{th}$, *MeV* | *γ-ray energy, $E_\gamma$/keV ($I_\gamma$/ %)* | *Half-life, $T_{1/2}$* |
|---|---|---|---|---|
| $^{73}$Se | $^{74}$Se(γ,1*n*) | 12.07 | 67.07 (70), 361.2 (97) | 7.15 h |
| $^{73m}$Se | $^{74}$Se(γ,1*n*) | 12.07 | 253.70 (2.36) | 39.8 m |
| $^{75}$Se | $^{76}$Se(γ,1*n*) | 11.15 | 121.12 (17.2), 136.00 (58.50), 264.66 (58.90), 279.54 (25.02), 400.66 (11.41) | 119.78 d |
| $^{81}$Se | $^{82}$Se(γ,1*n*) | 9.27 | 275.93 (0.68), 290.04 (0.56), 566.03 (0.224) | 18.45 m |
| $^{81m}$Se | $^{82}$Se(γ,1*n*) | 9.27 | 103.01 (12.8) | 57.28 m |
| $^{73}$As | $^{74}$Se(γ,1*p*) | 8.54 | 53.437 (10.6) | 80.3 d |
| $^{76}$As | $^{77}$Se(γ,1*p*) | 9.59 | 559.10 (45), 657.05 (6.2), 1216.08 (3.42) | 1.09 d |
| $^{77}$As | $^{78}$Se(γ,1*p*) | 10.39 | 239.01 (1.59), 249.81 (0.39), 520.65 (0.56) | 1.62 d |
| $^{79}$As | $^{80}$Se(γ,1*p*) | 11.41 | 95.73 (9.3), 365.0 (1.86), 432.1 (1.49) | 9.01 m |

## 3. Results and discussion

The experimental yields of the reactions $Y_{\text{exp}}$ are normalized to one electron of the accelerated beam incident on the bremsstrahlung target and calculated using the following formula:

$$Y_{exp} = \frac{S_p \cdot C_{abs}}{\varepsilon \cdot I_\gamma} \frac{t_{real}}{t_{live}} \frac{1}{N} \frac{1}{N_e} \frac{e^{\lambda \cdot t_{cool}}}{(1 - e^{-\lambda \cdot t_{real}})} \frac{\lambda \cdot t_{irr}}{(1 - e^{-\lambda \cdot t_{irr}})}, \quad (1)$$

where $S_p$ is the full-energy-peak area, $\varepsilon$ is the full-energy-peak detector efficiency, $I_\gamma$ is the gamma emission probability; $C_{abs}$ is the correction for self-absorption of gamma rays in the sample; $t_{real}$ and $t_{live}$ are the real time and the live time of the measurement, respectively; $N$ is the number of atoms in the activation sample, $N_e$ is the integral number of incident electrons, $\lambda$ is the decay constant, $t_{cool}$ is the cooling time, and $t_{irr}$ is the irradiation time.

The yields $Y_{\text{theor}}$ of photonuclear reactions representing the convolution of the photonuclear reactions cross section σ(*E*) and the distribution density of the number of bremsstrahlung photons over energy per one electron of the accelerator $W(E, E_{\gamma\text{max}})$ were determined as a result of the experiment. For the yield measurement for a natural mixture of isotopes the result is the yield of isotope production in all possible reactions on a natural mixture:

$$Y_{theor} = \sum_i \eta_i \int_{E_{th}}^{E_{\gamma max}} \sigma(E) W(E, E_{\gamma max}) dE \quad (2)$$

where $E_{\gamma\text{max}}$ is the kinetic energy of electrons hitting the tungsten radiator, *E* is the energy of bremsstrahlung photons produced on the radiator, $E_{th}$ is the threshold of the studied photonuclear reaction, η is the percentage of the studied isotope in the natural mixture of selenium isotopes, the index *i* corresponds to the number of the reaction contributing to the production of the studied isotope.

The total and partial cross sections σ(*E*) of the photonuclear reactions on the selenium isotopes were computed for the monochromatic photons with the TALYS1.96 code [22] with the standard parameters and CMPR [23]. The TALYS program analyzes all reactions occurring in the nucleus and transitions between states. Therefore, it is possible to determine not only the total cross sections of photonuclear reactions, but also the cross sections of reactions with the formation of specific states, in particular isomeric states. CMPR provides calculation of cross sections of photonuclear reactions with production of a studied isotope, i.e., the sum of the ground and isomeric states. For the yield measurement for a natural mixture of isotopes the result is the yield of isotope production in all possible reactions on a natural mixture. In our case, each radioactive nucleus was formed as a result of a specific one photonuclear reaction, since the thresholds of other formation channels exceeds 23 MeV.

The main disadvantage of beam experiments bremsstrahlung is that the yield of photonuclear reaction depends both on the studied cross section of the reaction σ(*E*) and the shape of the bremsstrahlung spectrum $W(E, E_{\gamma max})$, which is often known with insufficient accuracy. That's why the data obtained from the photonuclear experiment on bremsstrahlung beams are generally represented in terms of the relative yields or the integrated reaction cross section [33-35], the flux weighted average cross section <σ> [36-43], or the cross section per equivalent photon $\sigma_q$ [35, 37, 43, 44].

The use of the relative yields will make it possible to obtain the dependence of the probability of photonuclear reactions on the maximum energy of bremsstrahlung under different experimental conditions. The calibration with respect to the yield of the most probable reaction excludes the influence of the total photon absorption cross section. In our case, the dominant reaction is $^{82}Se(\gamma,1n)^{81m+g}Se$. Theoretical values of the relative yields can be calculated by the following formula:

$$Y_{rel,i} = \frac{\sum_i \eta_i \int_{E_{th}}^{E_{\gamma max}} \sigma_i(E) W(E, E_{\gamma max}) dE}{\eta_{Se-82} \int_{E_{th}}^{E_{\gamma max}} \sigma_{(\gamma,n)}(E) W(E, E_{\gamma max}) dE} \quad (3)$$

Due to the assumption on the unchanged shape of the bremsstrahlung spectrum, the bremsstrahlung photon production cross section $\sigma(E, E_{\gamma max})$ should be taken as the function $W(E, E_{\gamma max})$:

$$Y_{rel,i} = \frac{\sum_i \eta_i \int_{E_{th}}^{E_{\gamma max}} \sigma_i(E) \sigma(E, E_m) dE}{\eta_{Se-82} \int_{E_{th}}^{E_{\gamma max}} \sigma_{(\gamma,n)}(E) \sigma(E, E_m) dE} \quad (4)$$

where $\sigma(E, E_{\gamma max})$ is calculated based on the Zeltzer-Berger tables [45].

*Fig. 2 and Fig. 3 show e*xperimental values of the relative yields of photoneutron reactions normalized to the yield of the reaction $^{82}Se(\gamma,1n)^{81m+g}Se$ (Table 3 contains exactly this). In the case of the photoneutron reactions, theoretical calculations and experimental results are in good agreement with each other. Experimental values of relative yields lie closer to the curves of TALYS and CMPR.

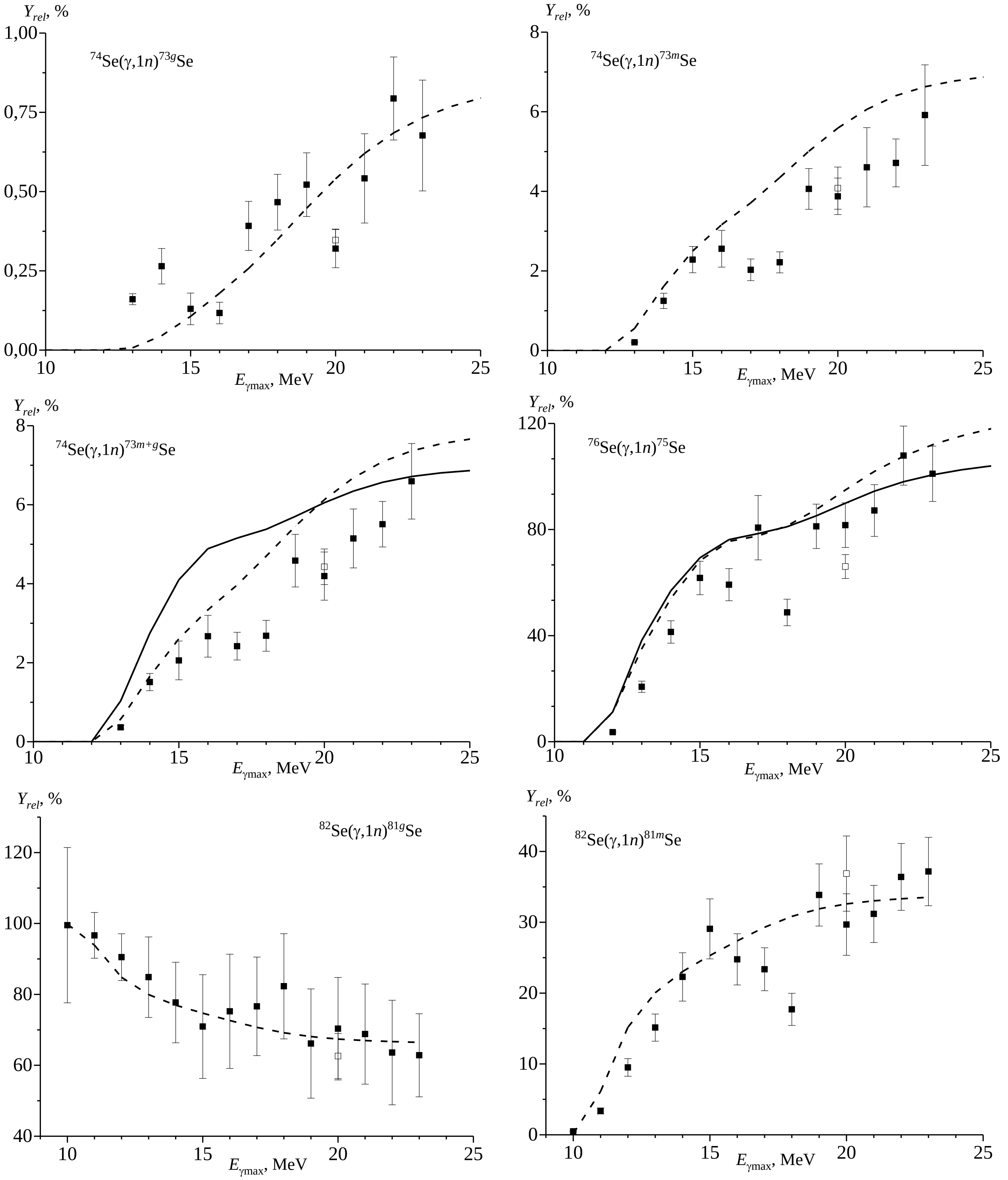

Figure 2. The relative yields of $^{nat}Se(\gamma,1n)$ reactions as a function of bremsstrahlung end-point energy from the present work (solid rectangles), literature data [14] (open rectangle) and the simulated values using CMPR (solid line) and TALYS code (dashed line) based on monoenergetic photons

Table 3. The relative yields of $^{nat}Se(\gamma,1n)$ reactions and comparison with theoretical results calculated on the basis of the TALYS and CMPR

| Reaction | $E_{\gamma max}$ | $Y_{rel}$, % | $Y_{relTALYS}$, % | $Y_{relCMPR}$, % |
|---|---|---|---|---|
| $^{74}Se(\gamma, n)^{73g}Se$ | *13 MeV* | 0.16 ± 0.02 | 0.01 | |
| | *14 MeV* | 0.26 ± 0.06 | 0.04 | |
| | *15 MeV* | 0.13 ± 0.05 | 0.10 | |
| | *16 MeV* | 0.12 ± 0.03 | 0.17 | |
| | *17 MeV* | 0.39 ± 0.08 | 0.25 | |
| | *18 MeV* | 0.47 ± 0.09 | 0.33 | |
| | *19 MeV* | 0.52 ± 0.10 | 0.43 | |
| | *20 MeV* | 0.32 ± 0.06<br>0.35 ± 0.03 [14] | 0.52 | |
| | *21 MeV* | 0.54 ± 0.14 | 0.59 | |
| | *22 MeV* | 0.79 ± 0.13 | 0.65 | |
| | *23 MeV* | 0.68 ± 0.17 | 0.70 | |
| $^{74}Se(\gamma, n)^{73m}Se$ | *13 MeV* | 0.20 ± 0.01 | 0.53 | |
| | *14 MeV* | 1.25 ± 0.19 | 1.54 | |
| | *15 MeV* | 1.93 ± 0.33 | 2.39 | |
| | *16 MeV* | 2.56 ± 0.46 | 3.02 | |
| | *17 MeV* | 2.03 ± 0.27 | 3.55 | |
| | *18 MeV* | 2.22 ± 0.26 | 4.16 | |
| | *19 MeV* | 4.06 ± 0.51 | 4.79 | |
| | *20 MeV* | 3.87 ± 0.46<br>4.08 ± 0.53 [14] | 5.35 | |
| | *21 MeV* | 4.60 ± 0.99 | 5.79 | |
| | *22 MeV* | 4.71 ± 0.60 | 6.12 | |
| | *23 MeV* | 5.92 ± 1.26 | 6.34 | |
| $^{74}Se(\gamma, n)^{73m+g}Se$ | *13 MeV* | 0.37 ± 0.05 | 0.54 | 1.03 |
| | *14 MeV* | 1.51 ± 0.22 | 1.58 | 2.75 |
| | *15 MeV* | 2.06 ± 0.49 | 2.49 | 4.10 |
| | *16 MeV* | 2.67 ± 0.53 | 3.19 | 4.89 |
| | *17 MeV* | 2.42 ± 0.350 | 3.79 | 5.15 |
| | *18 MeV* | 2.68 ± 0.39 | 4.49 | 5.38 |
| | *19 MeV* | 4.58 ± 0.67 | 5.22 | 5.70 |
| | *20 MeV* | 4.19 ± 0.61<br>4.43 ± 0.45 [14] | 5.86 | 6.05 |
| | *21 MeV* | 5.15 ± 0.75 | 6.39 | 6.35 |
| | *22 MeV* | 5.51 ± 0.58 | 6.78 | 6.57 |
| | *23 MeV* | 6.59 ± 0.96 | 7.04 | 6.71 |
| $^{76}Se(\gamma, n)^{75}Se$ | *12 MeV* | 3.60 ± 0.37 | 10.78 | 11.20 |
| | *13 MeV* | 20.73 ± 2.12 | 34.22 | 38.23 |
| | *14 MeV* | 41.38 ± 4.24 | 52.70 | 56.98 |
| | *15 MeV* | 61.73 ± 6.32 | 66.58 | 69.35 |
| | *16 MeV* | 59.20 ± 6.06 | 73.65 | 76.19 |
| | *17 MeV* | 80.70 ± 12.13 | 75.65 | 78.46 |
| | *18 MeV* | 48.74 ± 5.01 | 79.38 | 81.07 |
| | *19 MeV* | 81.20 ± 8.36 | 85.38 | 85.19 |
| | *20 MeV* | 81.65 ± 8.42<br>66.05 ± 4.51 [14] | 92.56 | 89.92 |
| | *21 MeV* | 87.18 ± 9.74 | 99.44 | 94.47 |
| | *22 MeV* | 107.88 ± 11.14 | 104.98 | 98.03 |
| | *23 MeV* | 101.02 ± 10.44 | 109.27 | 100.63 |
| $^{82}$Se(γ, *n*)$^{81g}$Se | *10 MeV* | 99.51 ± 21.89 | 99.84 | |
| | *11 MeV* | 96.64 ± 6.46 | 93.89 | |
| | *12 MeV* | 90.49 ± 6.59 | 84.83 | |
| | *13 MeV* | 84.86 ± 11.35 | 79.95 | |
| | *14 MeV* | 77.71 ± 11.36 | 76.95 | |
| | *15 MeV* | 70.93 ± 14.63 | 74.69 | |
| | *16 MeV* | 75.23 ± 16.11 | 72.63 | |
| | *17 MeV* | 76.63 ± 13.90 | 70.71 | |
| | *18 MeV* | 82.29 ± 14.84 | 69.17 | |
| | *19 MeV* | 66.14 ± 15.39 | 68.10 | |
| | *20 MeV* | 70.33 ± 14.45<br>62.59 ± 6.37 [14] | 67.41 | |
| | *21 MeV* | 68.82 ± 14.12 | 66.97 | |
| | *22 MeV* | 63.60 ± 14.75 | 66.68 | |
| | *23 MeV* | 62.84 ± 11.72 | 66.48 | |
| $^{82}$Se(γ, *n*)$^{81m}$Se | *10 MeV* | 0.49 ± 0.08 | 0.16 | |
| | *11 MeV* | 3.36 ± 0.43 | 6.11 | |
| | *12 MeV* | 9.51 ± 1.24 | 15.17 | |
| | *13 MeV* | 15.13 ± 1.92 | 20.04 | |
| | *14 MeV* | 22.29 ± 3.41 | 23.04 | |
| | *15 MeV* | 29.07 ± 4.24 | 25.30 | |
| | *16 MeV* | 24.76 ± 3.61 | 27.36 | |
| | *17 MeV* | 23.37 ± 3.03 | 29.29 | |
| | *18 MeV* | 17.71 ± 2.26 | 30.82 | |
| | *19 MeV* | 33.86 ± 4.39 | 31.89 | |
| | *20 MeV* | 29.67 ± 4.35<br>36.87 ± 5.31 [14] | 32.59 | |
| | *21 MeV* | 31.18 ± 4.04 | 33.03 | |
| | *22 MeV* | 36.39 ± 4.72 | 33.32 | |
| | *23 MeV* | 37.16 ± 4.82 | 33.52 | |

Fig. 3 and Table 4 show the experimental values of the relative yields for the photoproton reactions on natural mixture of selenium, and also, the data computed with the use of the codes TALYS and CMPR. In the case of the $^{74}$Se(γ,1*p*) reaction, theoretical calculations and experimental results are in good agreement with each other. In the case of relative yields for photoproton reactions on the heavy selenium isotopes the theoretical values calculated using CMPR are much larger than the TALYS results. For photoproton reactions on the isotopes of $^{77}$Se, $^{78}$Se, and $^{80}$Se the ratios of theoretical relative yields $Y_{\mathrm{relCMPR}} / Y_{\mathrm{relTALYS}}$ with increasing energy increase in the range of 2-5, 3-11 and 11-23, respectively. Experimentally obtained results lie closer to the theoretical curve according to CMPR. Including isospin splitting in the CMPR allows to describe experimental data on reactions with proton escape in energies range from 10 to 23 MeV. At the energy region above 25 MeV in addition to isospin splitting, quadrupole resonance, the overtone of the giant resonance, and the quasideuteron mechanism make a significant contribution to the cross sections [14].

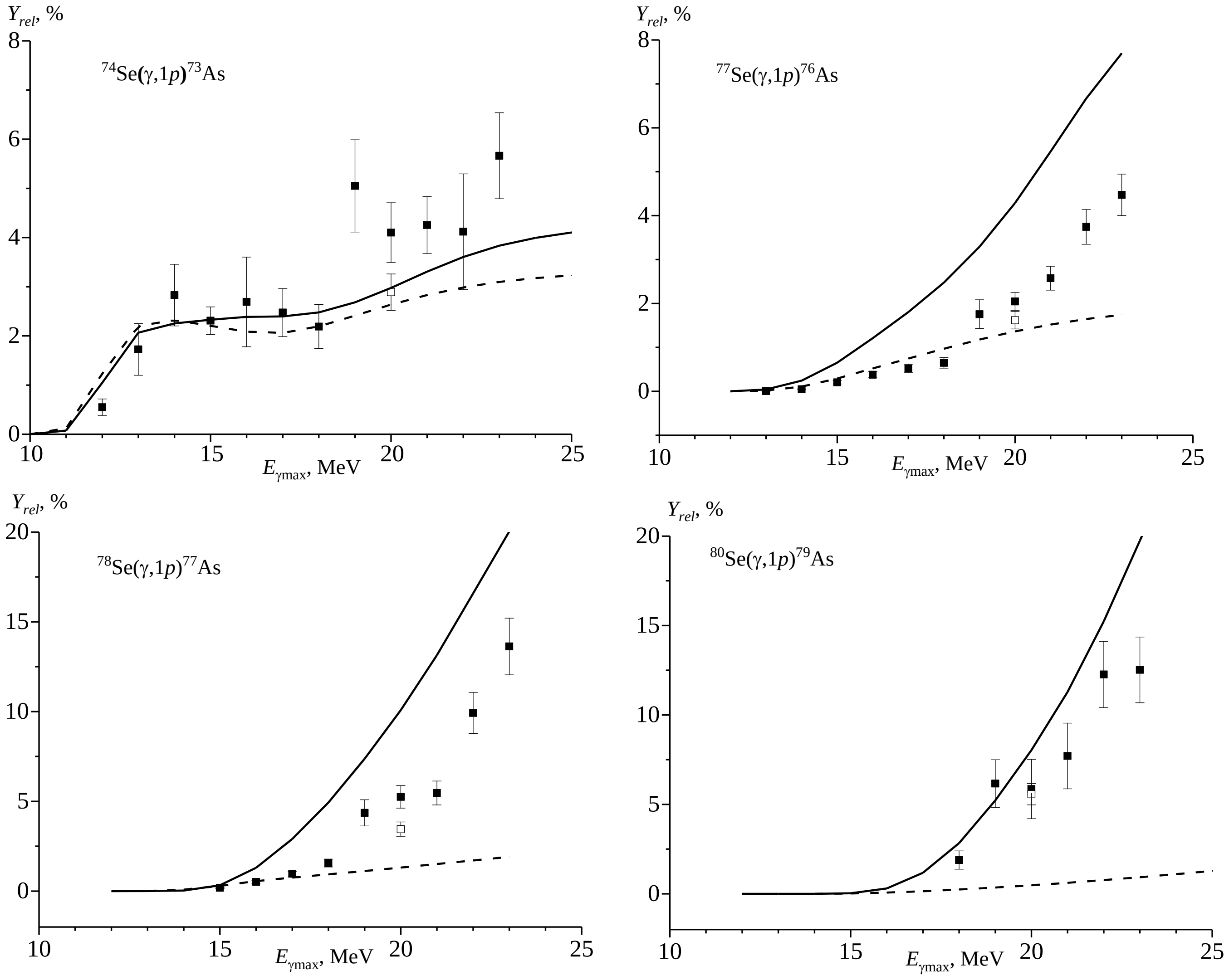

Figure 3. The relative yields of $^{nat}$Se(γ,1$p$) reactions as a function of bremsstrahlung end-point energy from the present work (solid rectangles), literature data [14] (open rectangle) and the simulated values using CMPR (solid line) and TALYS code (dashed line) based on monoenergetic photons

Table 4. The relative yields of $^{nat}$Se(γ,1$p$) reactions and comparison with theoretical results calculated on the basis of the TALYS and CMPR

| Reaction | $E_{\gamma max}$ | $Y_{rel}$, % | $Y_{relTALYS}$, % | $Y_{relCMPR}$, % |
|---|---|---|---|---|
| $^{74}$Se(γ, $p$)$^{73}$As | *12 MeV* | 0.55 ± 0.17 | 1.23 | 1.05 |
| | *13 MeV* | 1.72 ± 0.52 | 2.20 | 2.06 |
| | *14 MeV* | 2.83 ± 0.63 | 2.32 | 2.25 |
| | *15 MeV* | 2.31 ± 0.28 | 2.20 | 2.33 |
| | *16 MeV* | 2.69 ± 0.91 | 2.08 | 2.39 |
| | *17 MeV* | 2.47 ± 0.49 | 2.06 | 2.39 |
| | *18 MeV* | 2.19 ± 0.45 | 2.19 | 2.48 |
| | *19 MeV* | 5.05 ± 0.94 | 2.41 | 2.68 |
| | *20 MeV* | 4.10 ± 0.61<br>2.89 ± 0.37 [14] | 2.64 | 2.97 |
| | *21 MeV* | 4.25 ± 0.58 | 2.83 | 3.30 |
| | *22 MeV* | 4.12 ± 1.18 | 2.98 | 3.60 |
| | *23 MeV* | 5.66 ± 0.87 | 3.09 | 3.83 |
| $^{77}$Se(γ, *p*)$^{76}$As | *13 MeV* | 0.005 ± 0.001 | 0.0004 | 0.002 |
| | *14 MeV* | 0.05 ± 0.01 | 0.02 | 0.04 |
| | *15 MeV* | 0.21 ± 0.04 | 0.11 | 0.24 |
| | *16 MeV* | 0.38 ± 0.07 | 0.29 | 0.65 |
| | *17 MeV* | 0.52 ± 0.09 | 0.52 | 1.21 |
| | *18 MeV* | 0.65 ± 0.12 | 0.75 | 1.81 |
| | *19 MeV* | 1.76 ± 0.33 | 0.97 | 2.47 |
| | *20 MeV* | 2.05 ± 0.21<br>1.62 ± 0.20 [14] | 1.18 | 3.29 |
| | *21 MeV* | 2.57 ± 0.27 | 1.36 | 4.29 |
| | *22 MeV* | 3.74 ± 0.39 | 1.52 | 5.46 |
| | *23 MeV* | 4.47 ± 0.47 | 1.65 | 6.67 |
| $^{78}$Se(γ, *p*)$^{77}$As | *15 MeV* | 0.18 ± 0.03 | 0.29 | 0.32 |
| | *16 MeV* | 0.51 ± 0.07 | 0.56 | 1.29 |
| | *17 MeV* | 0.96 ± 0.13 | 0.76 | 2.90 |
| | *18 MeV* | 1.56 ± 0.22 | 0.93 | 4.94 |
| | *19 MeV* | 4.36 ± 0.73 | 1.12 | 7.37 |
| | *20 MeV* | 5.24 ± 0.63<br>3.45 ± 0.40 [14] | 1.31 | 10.08 |
| | *21 MeV* | 5.46 ± 0.66 | 1.51 | 13.14 |
| | *22 MeV* | 9.92 ± 1.14 | 1.71 | 16.59 |
| | *23 MeV* | 13.63 ± 1.57 | 1.91 | 20.04 |
| $^{80}$Se(γ, *p*)$^{79}$As | *18 MeV* | 1.89 ± 0.51 | 0.24 | 2.83 |
| | *19 MeV* | 6.16 ± 1.33 | 0.35 | 5.22 |
| | *20 MeV* | 5.86 ± 1.66<br>5.57 ± 0.60 [14] | 0.48 | 8.03 |
| | *21 MeV* | 7.71 ± 1.84 | 0.61 | 11.29 |
| | *22 MeV* | 12.27 ± 1.85 | 0.77 | 15.23 |
| | *23 MeV* | 12.52 ± 1.84 | 0.93 | 19.75 |

Unlike such widely spread numerical codes as TALYS, GNASH, and EMPIRE, the CMPR considers not only the giant dipole resonance and the quasideuteron photoabsorption mechanism, but also the contribution into the cross section of isovector quadrupole resonance and the GDR overtone in calculation of the photoabsorption cross section. The energies and integral cross sections of this giant resonances are calculated in the framework of the semimicroscopic model with multipole–multipole residual forces [23].

In nuclei with N ≠ Z, upon absorption of electric dipole γ photons, two branches of the giant dipole resonance are excited, $T_{<} = T_0$ and $T_{>} = T_0 + 1$, where $T_0 = \frac{|N-Z|}{2}$. Fig. 4 shows the excitations of the isospin components $T_<$ and $T_>$ of the giant dipole resonance in initial nucleus ($N$, $Z$) and their decay according to the proton ($N$, $Z - 1$) and neutron ($N - 1$, $Z$) channels. From Fig. 4 it can be seen that the decay of excited GDR states with isospin $T_{>} = T_0 + 1$ according to the neutron channel to low-lying states $T = T_0 - 1/2$ with neutron emission is forbidden, which leads to increase in the reaction cross section (γ, 1*p*) and to a shift maximum of the reaction cross section (γ, 1*p*) with respect to reactions (γ, 1*n*) towards higher energies in the nucleus (*N, Z*).

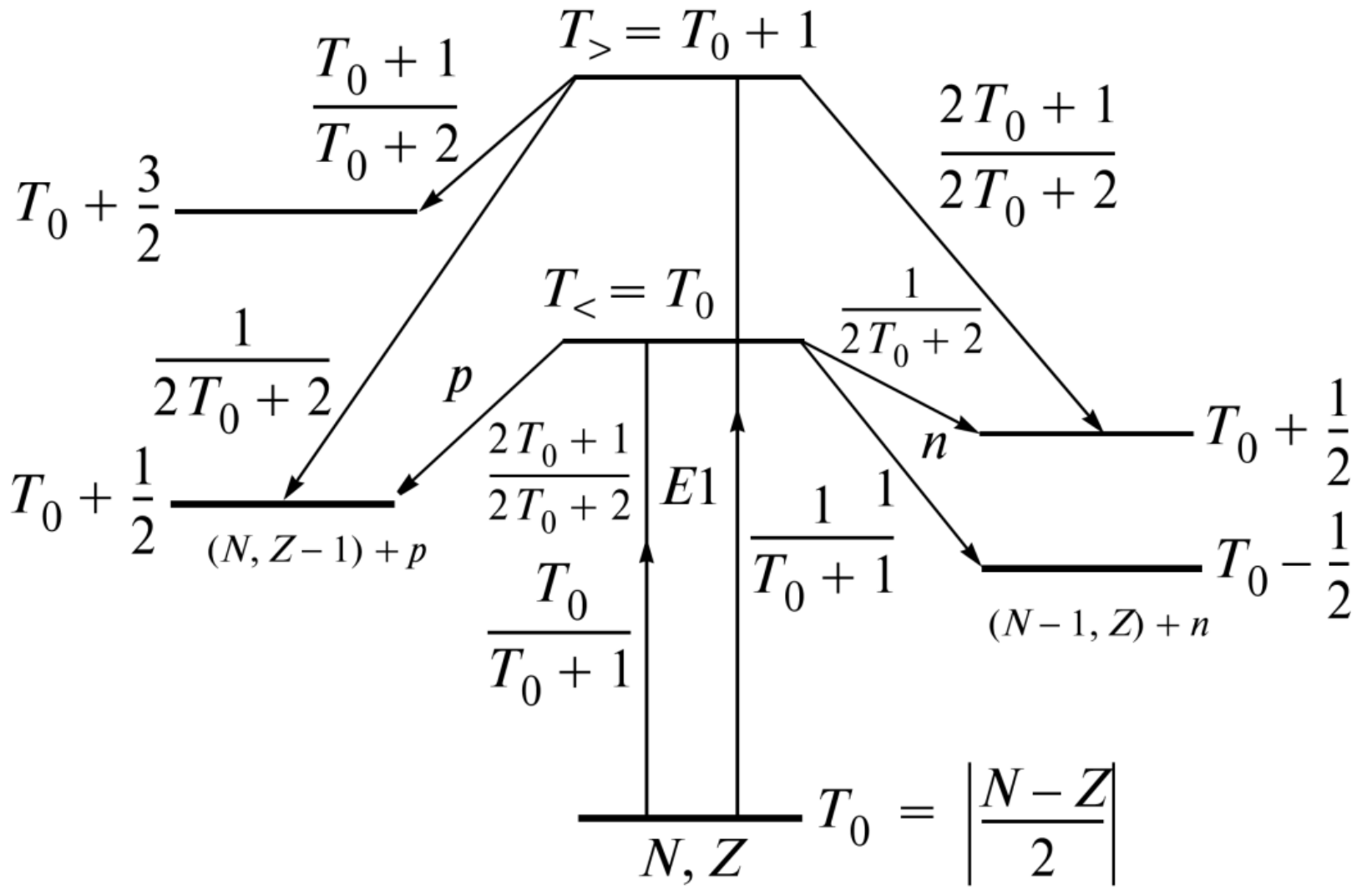

Figure 4. Scheme of excitation of states $T_<$ and $T_>$ in the nucleus (*N, Z*) and their decay along the proton channel ($N, Z - 1$) and the neutron channel ($N - 1, Z$)

The value of isospin splitting of GDR is determined by the relation (5):

$$\Delta E = E(T_>) - E(T_<) = \frac{60}{A}(T_0 + 1) \tag{5}$$

For isotopes $^{74,77,78,80}$Se, the isospin increases from 3 to 6, which leads to an increase in the isospin splitting of GDR for these isotopes from 3.24 to 5.25 MeV.

The ratio of the probabilities of excitation of states T> and T< is described by the relation (6):

$$\frac{s(T_>)}{s(T_<)} = \frac{1}{T_0}\frac{1 - 1.5T_0A^{-2/3}}{1 + 1.5T_0A^{-2/3}} \tag{6}$$

For isotopes $^{74,77,78,80}$Se, the ratio $s(T_>)/s(T_<)$ decreases from 0.20 to 0.06 with an increase in the mass number A. Thus, for isotopes $^{74,77,78,80}$Se, the isospin splitting of GDR increases with an increase in the mass number A, but the relative role of the excitation channel decreases.

The decay of excited GDR states with isospin $T_> = T_0 + 1$ according to the neutron channel to low-lying states $T = T_0 - 1/2$ with neutron emission is forbidden, which leads to increase in the reaction cross section (γ, 1*p*) and to a shift maximum of the reaction cross section (γ, 1*p*) with respect to reactions (γ, 1*n*) towards higher energies in the nucleus (*N, Z*). Fig. 5 and Fig. 6 show contribution of the $T_<$- and $T_>$-components to the theoretical cross sections and the relative yields for photoproton reactions of $^{74,77,78,80}$Se isotopes. As can be seen on Fig. 5 in heavy isotopes of selenium, isospin splitting plays a significant role, the accounting of which makes it possible to correctly describe GDR decay photoproton channel. The experimental data obtained by us also confirm this fact.

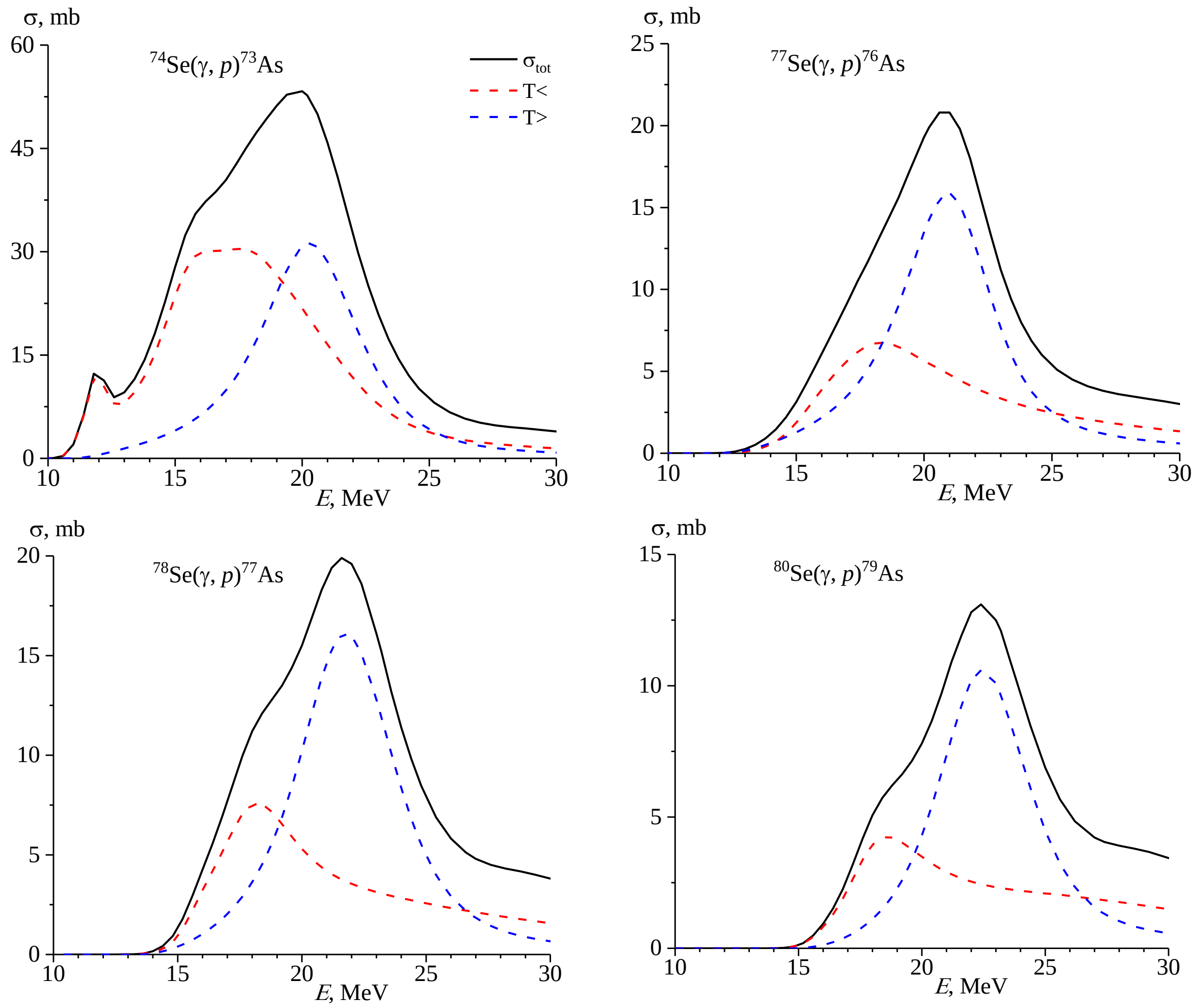

Figure 5. The cross section reactions and the cross sections of the GDR components $T_< = T_0$ and $T_>=T_0 +1$ for reaction (γ, 1$p$) on $^{74,77,78,80}$Se isotopes

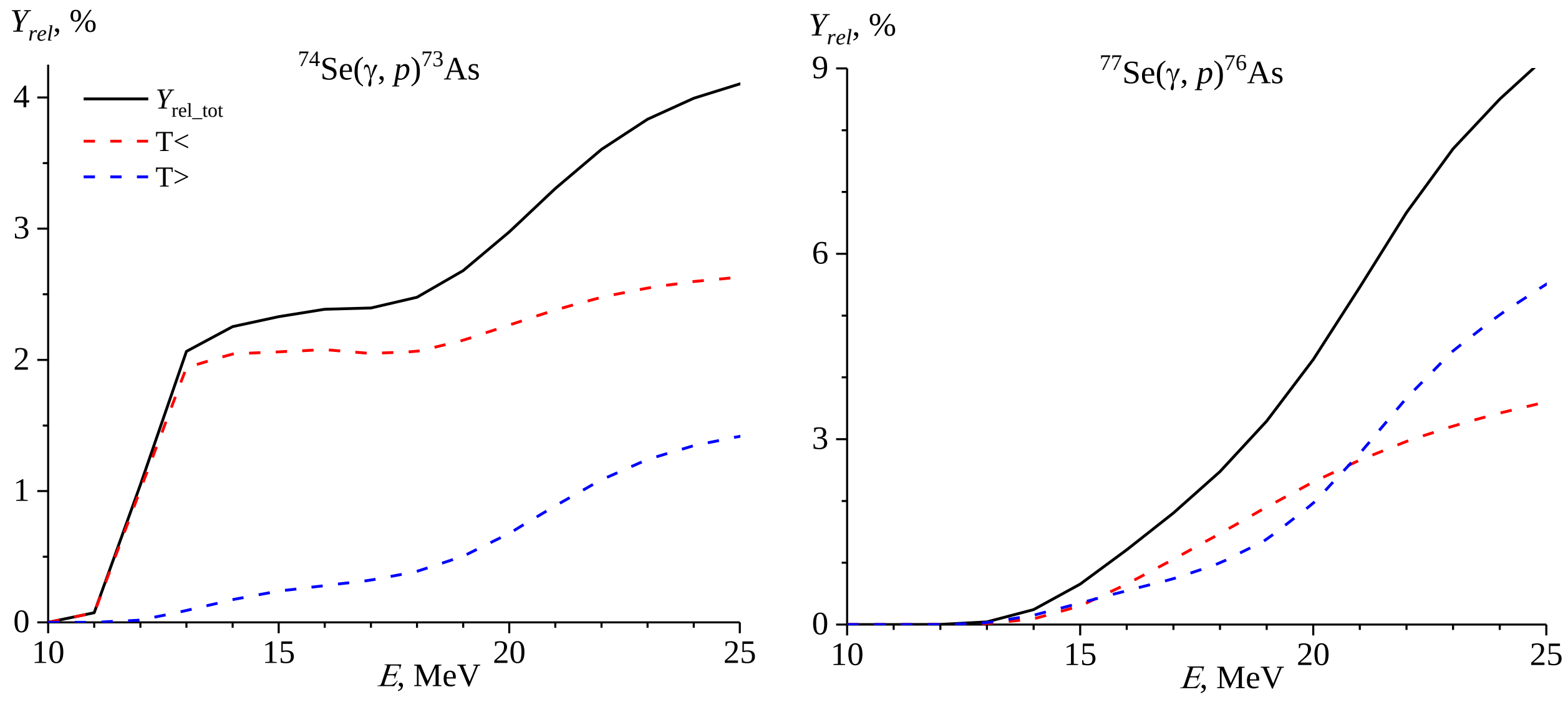

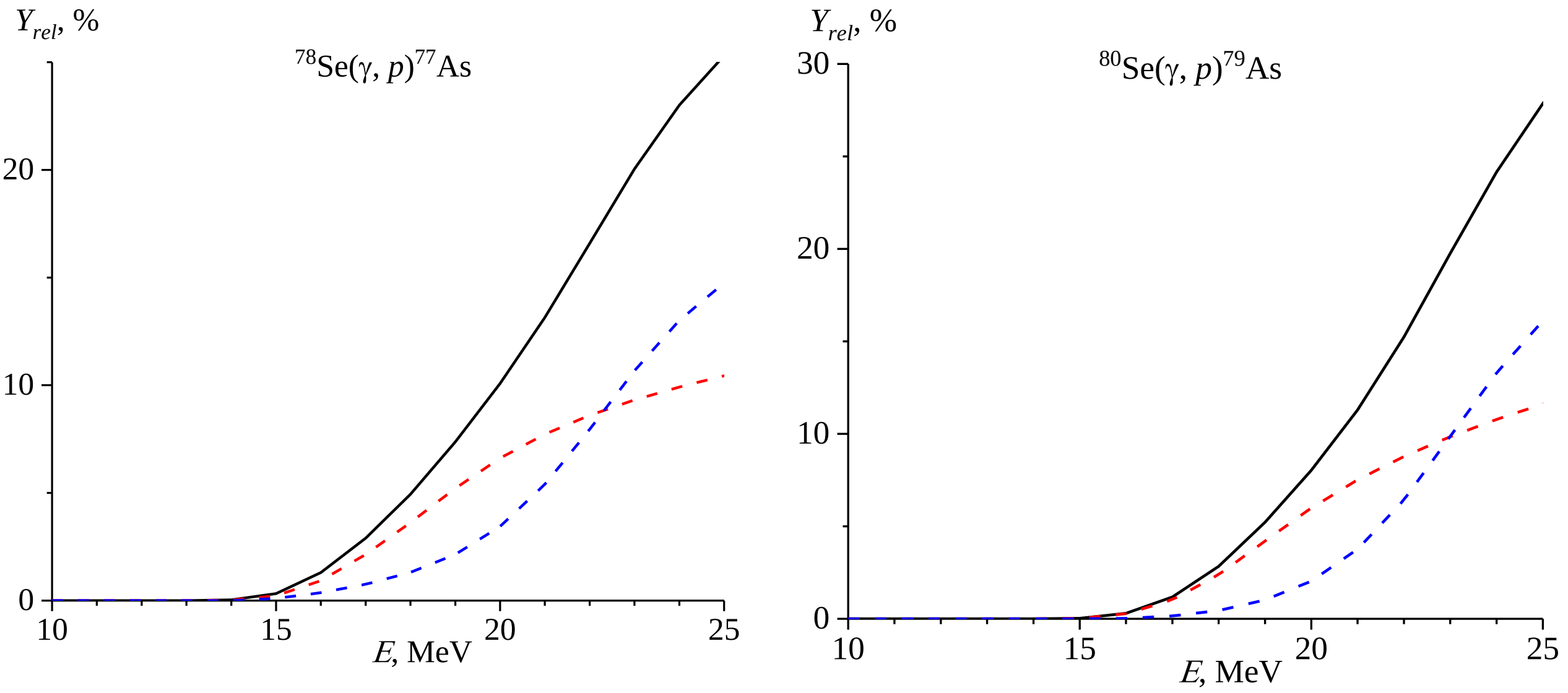

Figure 6. Contribution of the $T_<$ and $T_>$ components to the theoretical relative yields for photoproton reactions on a natural mixture of selenium isotopes

Among the nuclei heavier than $^{56}$Fe, There is a group of 35 neutron-deficient stable nuclei, starting with $^{74}$Se, with a very low abundance in solar sistem, the formation of which cannot be described by neutron capture reactions. Many production sites of the p-nuclei have been proposed: oxygen/neon layers of highly evolved massive stars during their presupernova phase [46] and during their supernova explosion [47], X-ray novae [48], neutrinodriven winds originating from a nascent neutron star shortly after supernova explosion [49], Type Ia supernova explosions [50], and helium-accreting CO white dwarfs of sub-Chandrasekhar mass [51]. Among them, the most promising is the oxygen/neon layers during a Type II supernova explosion. The p-nuclei are synthesized by the photodisintegration of s-nuclei (*s*-process seeds) produced in the layers during the core helium burning in the progenitor. The production of p-nuclei via the subsequent photodisintegration is referred to as a p-process [52]. Photonuclear reactions  are threshold, so a necessary condition for their occurrence is a high temperature T = 1 - 3.5 K [47], which is fulfilled when a shock wave passes through the layers of a pre-supernova star of the SnII type after the collapse of the supernova core [53-54]. For the synthesis of *p*-nuclei, in addition to the considered gamma-ray processes on equilibrium photons, other models have been proposed: for example, nuclear reactions of proton capture (*p*,γ), (*p*,*n*), reactions under the influence of powerful fluxes of neutrino radiation from the stellar core, and also reactions of rapid capture of protons accreted on the surface of a neutron star [55]. To describe the formation and decay of *p*-nuclei as a result of photonuclear reactions, it is necessary to accurately know the yields of photoproton and photoneutron reactions, the correct calculation of which is impossible without taking into account the isospin splitting of GDR. The *p*-process involves positron production and capture, proton capture, and (γ, *n*) or (*p*, *n*) reactions starting from the *s*- and *r*-isotopes as seed nuclei. Fig. 7 shows paths of the formation and decay of $^{74}$Se *p*-nuclide in stellar nucleosynthesis. Fig. 8*a* shows the cross sections of (γ,1*n*), (γ,2*n*) and (γ,3*n*) reactions corresponding to the isotopes $^{75}$Se, $^{76}$Se and $^{77}$Se calculated based on the CMPR. From the data shown in Fig. 8*a*, it follows that the main reactions of formation of the isotope $^{74}$Se will be the reactions ($\gamma$,1$n$) and ($\gamma$,2$n$). This leads to a significant buildup of the $p$-nucleus, assisted by a moderately strong $^{75}$As($\gamma$, $n$)$^{74}$As($\beta^-$)$^{74}$Se branch.

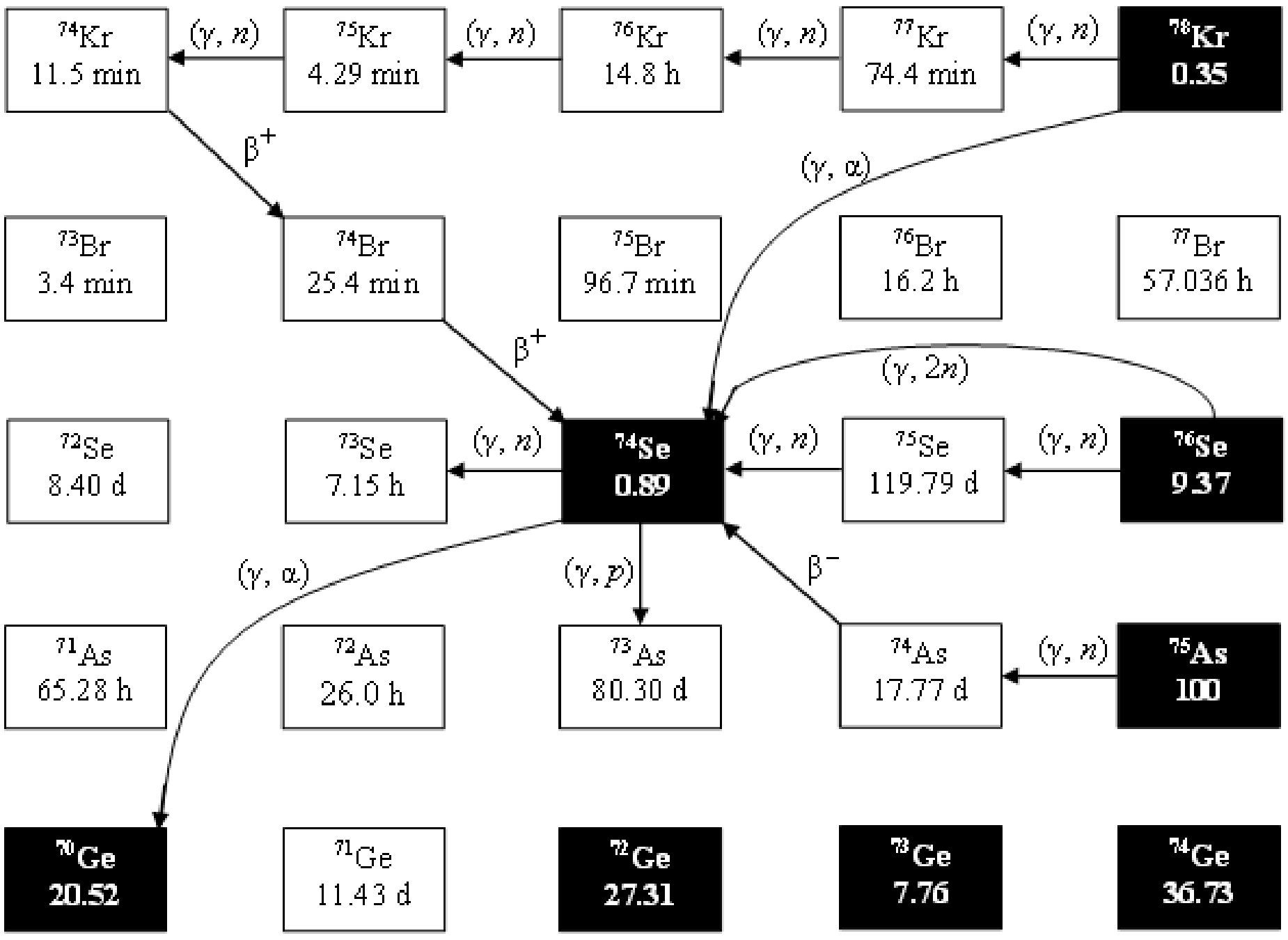

Figure 7. Paths of the production and decay of $^{74}$Se $p$-nuclide in stellar nucleosynthesis

As shown on Fig 7. there are three competing pathways for the decay of $^{74}$Se: ($\gamma$,1$n$) (12.07 MeV), ($\gamma$,1$p$) (8.54 MeV) and ($\gamma$,$\alpha$) (4.07 MeV). $^{74}$Se has a two-neutron separation energy (20.46 MeV) greater than 20 MeV, and thus cannot be destroyed via ($\gamma$, 2$n$) reaction. To compare the main destruction channels of the bypassed nucleus the cross sections of ($\gamma$,$n$), ($\gamma$,$p$) and ($\gamma$,$\alpha$) reactions on the isotope $^{74}$Se calculated based on the TALYS are given in Fig. 8$b$. The calculated reaction cross sections at maximum GDR are 100, 35 and 3 mb, respectively. The results of our research will allow us to experimentally compare $^{74}$Se($\gamma$,1$n$) and $^{74}$Se($\gamma$,1$p$) reactions. Fig 8$c$. shows the relative yields of $^{74}$Se($\gamma$,1$n$) and $^{74}$Se($\gamma$,1$p$) reactions. It can be seen from Fig 8$c$. that the two main competing ways of decay of the $^{74}$Se are almost equal. The reaction product $^{74}$Se($\gamma$,$\alpha$)$^{70}$Ge is stable and it is impossible to estimate the probability of passing this reaction using the gamma activation method.

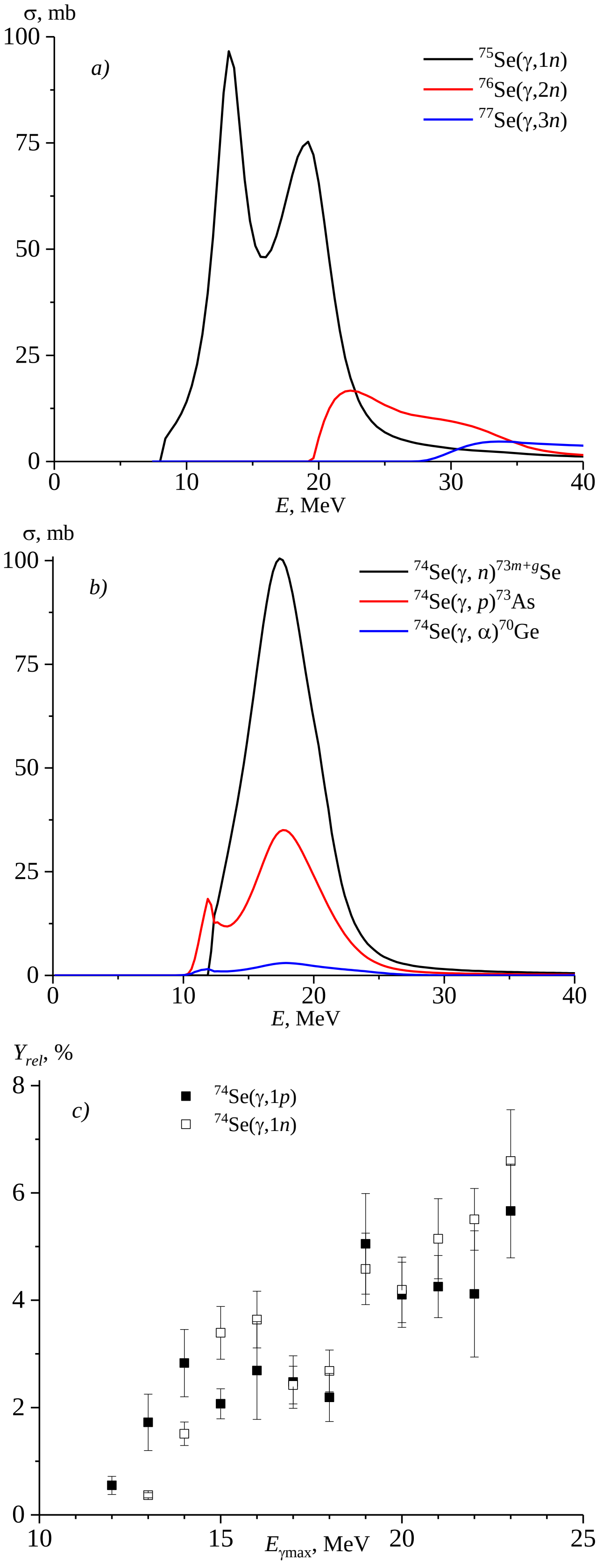

Figure 8. (*a*) The cross sections of (γ,1*n*), (γ,2*n*) and (γ,3*n*) reactions corresponding to the isotopes $^{75}$Se, $^{76}$Se and $^{77}$Se calculated based on the TALYS; (*b*) The cross sections of (γ,*n*), (γ,*p*) and (γ,α) reactions on the isotope $^{74}$Se calculated based on the TALYS; (*c*) The relative yields of reactions with the emission of a neutron (open rectangles) and a proton (solid rectangles) as a function of bremsstrahlung end-point energy

## 4. Conclusion

The present work has been concerned with measurements of the relative yields for the photonuclear reactions on natural mixture of selenium using bremsstrahlung end-point energies of 10 to 23 MeV. The bremsstrahlung photon flux was computed in the Geant4.11.1 code. Experimental results were compared to calculations using the TALYS model with the standard parameters and the CMPR. For the obtained photoneutron reactions, there is a good agreement between the experimental relative yields and calculations according to the TALYS program and in the CMPR framework. For the photoproton reaction on the light isotope $^{74}$Se, there is no difference between the data calculated using TALYS and CMPR and the experimental values. On the heavy selenium isotopes, the theoretical relative yields calculated using CMPR are much larger than the TALYS results. Including isospin splitting in the CMPR allows to describe experimental data on reactions with proton escape in energies range from 10 to 23 MeV. Therefore, taking into account isospin splitting is necessary for a correct description of the decay of the GDR. At the energy region above 25 MeV in addition to isospin splitting, quadrupole resonance, the overtone of the giant resonance, and the quasideuteron mechanism make a significant contribution to the cross sections. Study of photonuclear reactions on selenium isotopes is important for understanding the formation and decay of bypassed nuclei during nucleosynthesis.

**Acknowledgment**

The authors would like to thank the staff of the MT-25 microtron of the Flerov Laboratory of Nuclear Reactions, Joint Institute for Nuclear Research for their cooperation in the realization of the experiments.